\documentclass[sigconf,authorversion]{acmart}
\usepackage{amsmath}
\usepackage{natbib}
\usepackage{amsfonts}
\usepackage{subcaption}
\usepackage{hhline}
\usepackage{hyperref}
\usepackage{enumerate}
\usepackage{multirow}
\usepackage{array}
\usepackage[para]{footmisc}
\usepackage{bbm}
\usepackage{adjustbox}
\usepackage[algo2e,ruled,vlined]{algorithm2e}

\DeclareMathOperator*{\argmax}{arg\,max}

\usepackage{subcaption}

\usepackage[show]{chato-notes}
\newcommand{\iadh}[1]{\textcolor{black}{#1}}
\newcommand{\craig}[1]{\textcolor{black}{#1}}

\newcommand{\xiao}[1]{\textcolor{black}{#1}}
\newcommand{\xiaow}[1]{\textcolor{black}{#1}}

\newcommand{\nic}[1]{\textcolor{black}{#1}}
\newcommand{\craigbeta}[1]{\textcolor{black}{#1}}
\newcommand{\craigd}[1]{\textcolor{black}{#1}}

\newcommand{\craigi}[1]{\textcolor{black}{#1}}
\newcommand{\xiaoi}[1]{\textcolor{black}{#1}}
\newcommand{\io}[1]{\textcolor{black}{#1}}
\newcommand{\yaya}[1]{\textcolor{black}{#1}}

\newcommand{\craigj}[1]{\textcolor{black}{#1}}
\newcommand{\nicj}[1]{\textcolor{black}{#1}}
\newcommand{\xiaoj}[1]{\textcolor{black}{#1}}

\newcommand{\xiaoc}[1]{\textcolor{black}{#1}}
\newcommand{\craigc}[1]{\textcolor{black}{#1}}

\def\expcent{f_e}
\def\expdoc{f_b}
\def\clusters{K}
\def\centroid{\upsilon}
\def\IDF{\sigma}

\def\fbweight{\beta}
\def\emb{\phi}
\def\Fbembs{\Phi}

\usepackage{marginnote}
\newcommand{\pageenlarge}[1]{\enlargethispage{#1\baselineskip}}

\copyrightyear{2021} 
\acmYear{2021} 
\setcopyright{acmcopyright}\acmConference[ICTIR '21]{Proceedings of the 2021 ACM SIGIR International Conference on the Theory of Information Retrieval}{July 11, 2021}{Virtual Event, Canada}
\acmBooktitle{Proceedings of the 2021 ACM SIGIR International Conference on the Theory of Information Retrieval (ICTIR '21), July 11, 2021, Virtual Event, Canada}
\acmPrice{15.00}
\acmDOI{10.1145/3471158.3472250}
\acmISBN{978-1-4503-8611-1/21/07}

\begin{document}

\title[\iadh{Pseudo-Relevance} Feedback for Multiple Representation Dense Retrieval]{Pseudo-Relevance Feedback for\\ Multiple  Representation Dense Retrieval}

\def\name{\textsf{ColBERT-PRF}}
\def\colbertauths{Khattab \& Zaharia}

\author[X. Wang et al.]{Xiao Wang$^1$, Craig Macdonald$^1$, Nicola Tonellotto$^2$, Iadh Ounis$^1$}
\affiliation{%
  \institution{$^1$University of Glasgow, $^2$University of Pisa}
  \country{UK, Italy}
}
\renewcommand{\authors}{Xiao Wang,  Craig Macdonald, Nicola Tonellotto, Iadh Ounis}

\begin{abstract}
\looseness-1 Pseudo-relevance feedback mechanisms, from Rocchio to \iadh{the} relevance models, have \iadh{shown} the usefulness of expanding and reweighting \iadh{the} users' initial queries using information occurring in an initial set of retrieved documents, known as the pseudo-relevant set. 
\nicj{Recently, dense retrieval -- through the use of neural contextual language models such as BERT for analysing the documents' and queries' contents and computing their relevance scores -- has shown \yaya{a promising} performance on several information retrieval tasks still relying on the traditional inverted index for identifying documents relevant to a query.}
Two different dense retrieval \craigbeta{families} have emerged: the use of single embedded representations for each passage and query (e.g. using BERT's [CLS] token), or via multiple representations (e.g. using an embedding for each token of the query and document). 
In this work, we \io{conduct} \xiao{the} first study into the potential for multiple representation dense retrieval to be enhanced using pseudo-relevance feedback. In particular, based on the pseudo-relevant set of documents identified using a first-pass dense retrieval, we extract representative feedback embeddings \craigc{(using KMeans clustering)} -- while ensuring that these embeddings \craig{discriminate among passages} \craigc{(based on IDF)} -- which are then added to the query representation. These additional \craig{feedback} embeddings are shown to both enhance \craig{the effectiveness of} a reranking as well as an additional dense retrieval operation. \xiaow{Indeed, experiments \craigbeta{on the MSMARCO passage ranking dataset} show  that MAP can be improved by upto \xiaoi{26\%} on \io{the} TREC 2019 \io{query set} and \xiaoi{10\%} on \io{the} TREC 2020 \io{query set} \iadh{by the} application of our proposed \name{} \iadh{method on} a ColBERT dense retrieval approach.} 
\end{abstract}


\maketitle

\pageenlarge{0}\section{Introduction}





\craig{Query expansion approaches, which rewrite the user's query, have been shown} to be an effective approach to alleviate the \iadh{vocabulary} discrepancies between the user query and the \craig{relevant} documents, by \craig{modifying} the \iadh{user's} original query to improve the retrieval effectiveness. Many approaches follow the pseudo-relevance feedback (PRF) paradigm -- such as Rocchio's algorithm~\cite{rocchio1971relevance}, the RM3 relevance language model~\cite{abdulumass}, or the DFR query expansion models~\cite{amati2002probabilistic} -- where terms appearing in the top-ranked documents for the initial query are used to expand \iadh{it}. 
\craig{Query expansion \io{(QE)} approaches have also found a \iadh{useful} role when integrated} with effective BERT-based neural reranking models, by \craig{providing a high quality set of candidate documents obtained using the expanded query, which can then be reranked~\cite{padaki2020rethinking,yu2021pgt,wang2020pseudo}.} 

On the other hand, many studies have focused on the use of {\em static} word embeddings, such as \nic{\textsf{Word2Vec}}, \craig{within} query expansion methods~\cite{roy2016using,roy2018using,kuzi2016query,diaz2016query}. \io{Indeed}, most of the existing embedding-based QE methods~\cite{roy2016using,roy2018using,kuzi2016query,diaz2016query,zamani2016embedding} are based on static embeddings, where a word embedding is always the same within different \iadh{sentences}, \craig{and hence \iadh{they} do not address contextualised language models such as BERT.} \xiaoi{Recently, CEQE \cite{naseri2021ceqe} \craigi{was} proposed, which makes use of contextualised BERT embeddings for query expansion.} \craigi{The resulting refined query representation is then used for a further round of retrieval using a conventional inverted index.}
\xiaoi{In contrast, in this paper, we focus on implementing contextualised embedding-based query expansion for dense retrieval.}

\looseness -1 \craig{\xiaoi{Indeed,} \io{the} BERT models have demonstrated further promise in being a suitable basis for {\em dense retrieval}. In particular, instead of using a classical inverted index, in dense retrieval, \io{the} documents and queries are represented using embeddings. Then, \io{the} documents can be retrieved using an approximate nearest neighbour algorithm -- as exemplified by the FAISS toolkit}~\cite{JDH17}. Two distinct families of approaches have emerged: \xiao{single} representation dense retrieval and multiple representation dense retrieval. \nic{In single-representation dense retrieval, as used by DPR~\cite{karpukhin2020dense} and ANCE~\cite{xiong2020approximate}, each query \iadh{or} document is represented entirely by the single embedding of the [CLS] (classification) token computed by BERT.} \craig{Query-document relevance is estimated in terms of the similarity of the corresponding [CLS] embeddings.} In contrast, \iadh{in} multiple representation dense retrieval -- \craigi{as proposed by} ColBERT~\cite{khattab2020colbert} -- each term of the queries and documents is represented by a single embedding. For each query embedding\nic{, one per query term,} the nearest document token embeddings are identified using an \iadh{approximate} nearest neighbour search, before a final re-scoring to obtain exact relevance estimations. 


\looseness -1 In this work, we are concerned with applying pseudo-relevance feedback \iadh{in a} multiple representation dense retrieval setting. Indeed, as retrieval uses multiple representations, this allows additional \xiao{useful}  embeddings to be appended to the query representation. Furthermore, the exact scoring stage provides the document embeddings \xiao{in response to the original query,} which can be used as pseudo-relevance information. 
%
%
%
%
\xiao{Thus, in this work, we propose}
a pseudo-relevance feedback mechanism \xiao{called \name{}} for dense retrieval. \craigi{In particular, as embeddings cannot be counted, \name{} applies clustering to the embeddings occurring in the pseudo-relevant set, and then identifies the most discriminative embeddings among the cluster centroids. These centroids are then appended to the embeddings of the original query.} \craigi{\name{} is focussed on multiple representation dense retrieval settings; However, compared to existing work, our approach is the first work to apply pseudo-relevance feedback to any form of dense retrieval setting};
moreover, among the existing approaches applying deep learning for \iadh{pseudo-relevance} feedback, our \iadh{work in this paper} is the first that can improve the recall of the candidate set by re-executing the expanded query representation upon the dense retrieval index, \craigi{and thereby identify more relevant documents that can be highly ranked for the user.} 
\looseness -1 In summary, our work makes \xiaow{the following} contributions: (1) \xiaow{we propose a novel \craigj{contextualised} pseudo-relevance feedback mechanism for multiple representation dense retrieval;}
(2) we \craigbeta{cluster and rank} the feedback document embeddings for selecting candidate expansion embedding\nic{s}; (3) we evaluate our proposed \craigj{contextualised} \craig{PRF} model in \craig{both} ranking and reranking settings \craigc{and discuss its efficiency}. 


The remainder of this paper is as follows: \craigi{Section~\ref{sec:related} positions this work among existing approaches to pseudo-relevance feedback;} Section~\ref{sec:DenseRetrieval} describes a multi-representation dense retrieval, while Section~\ref{sec:DenseQE} presents our proposed dense PRF method. Next, we discuss our experimental setup and results in Section~\ref{sec:expsetup} \& \ref{sec:results}, respectively; We \iadh{provide concluding remarks in} Section~\ref{sec:conclusion}.


\section{Related Work}\label{sec:related}

Pseudo-relevance feedback approaches have a long history in Information Retrieval (IR) going back to Rocchio~\cite{rocchio1971relevance} who generated refined query reformulations through linear combinations of the sparse vectors (e.g. containing term frequency information) representing the query and the top-ranked feedback documents. 
Refined \nic{classical} PRF models, such as Divergence from Randomness's Bo1~\cite{amati2002probabilistic}, KL~\cite{amati2003probability}, and RM3 relevance models~\cite{abdulumass} have demonstrated their effectiveness on many test collections. Typically, these models identify and weight feedback terms that are frequent in the feedback documents and infrequent in the corpus, by exploiting statistical information about the \io{occurrence} of terms in \io{the} documents and in the whole collection.

Recently, deep learning solutions based on transformer networks have been used to enrich the statistical information about terms by rewriting or expanding the collection \io{of} documents. For instance, DeepCT~\cite{dai2020context} reweights terms occurring in \io{the} documents according to a fine-tuned  BERT model to \io{highlight} important terms. This results in {\em augmented} document representations, which can be indexed using a traditional inverted indexer. Similarly, doc2query~\cite{nogueira2019document} and its more modern variant docT5query~\cite{nogueira2019doc2query} apply text-to-text translation models to each document in the collection to suggest queries that may be relevant to the document. When \xiaoi{the suggested queries are} indexed along with the original document, \io{the retrieval} effectiveness is enhanced.

\nic{More recently, instead of leveraging (augmented) statistical information such as \io{the} in-document and collection frequency of terms to model a query or a document, dense representations, also known as embeddings, are becoming commonplace. Embeddings encode terms in queries and documents by learning a vector representation \io{for} each term, which takes into account the word semantic and context.}
Instead of identifying the related terms in the pseudo-relevance feedback documents using statistical methods, embedding-based query expansion methods~\cite{roy2016using,roy2018using,kuzi2016query,diaz2016query,zamani2016embedding} expand a query with terms that are closest to the query terms in the word embedding space. 
However, the expansion terms may not be sufficiently informative to distinguish relevant documents from non-relevant documents -- for instance, the embedding of ``grows'' may be closest to ``grow'' in the 
embedding space, but adding ``grows'' to the query may not help to identify more relevant documents. Moreover, all these embedding-based  method are based on non-contextualised embeddings, where a word embedding is always the same within different sentences, and hence they do not address contextualised language models.
\nic{\io{Pre-trained} contextualised language models such as BERT~\cite{devlin2019bert} have brought large effectiveness improvements over prior art \io{in} information retrieval tasks. In particular, deep learning is able to successfully exploit general language features in order to capture the contextual semantic signals allowing to better estimate the relevance of documents w.r.t. a given query.} 



\craigi{Query expansion approaches have been used for generating a high quality pool of candidate documents to be reranked by effective BERT-based neural reranking models~\cite{padaki2020rethinking,yu2021pgt,wang2020pseudo}.} \nic{However the use of BERT models directly within the pseudo-relevance feedback \io{mechanism} has seen comparatively little use in the literature. The current approaches leveraging \io{the} BERT contextualised embeddings for PRF are Neural PRF~\cite{li2018nprf}, BERT-QE~\cite{zheng2020bert} and CEQE~\cite{naseri2021ceqe}.}


\looseness -1 \nic{In particular, Neural PRF uses neural ranking models, such as DRMM~\cite{guo2016deep} and KNRM~\cite{xiong2017end}, to score the similarity of a document to a top-ranked feedback document. BERT-QE is conceptually similar to Neural PRF, but it measures the similarity of each document w.r.t. feedback chunks \io{that} are extracted from the top-ranked feedback documents. This results in an expensive application of many BERT computations -- approximately \craig{11$\times$ as many GPU operations than a simple BERT reranker}~\cite{zheng2020bert}.} \nic{\craig{Both Neural PRF and BERT-QE} approaches leverage contextualised language models to rerank \craig{an initial ranking} of documents retrieved by a preliminary sparse retrieval system. However, they cannot identify any new relevant documents from the collection that were not retrieved in the initial ranking.}

\nic{CEQE exploits BERT to compute contextualised representations \io{for} the query \io{as well as for} the terms in the top-ranked feedback documents, and then \io{selects} 
as expansion terms those 
which are the closest to the query embedding\xiaoi{s} according to some similarity measure. \craigi{In contrast to  Neural PRF and BERT-QE}, CEQE is used to generate a new query of terms for execution upon a conventional inverted index.} This means that the contextual meaning of an expansion term is lost - for instance, a polysemous word added to the query can result in \io{a topic} drift.
 
\looseness -1 \craigi{In contrast to the aforementioned approaches, our proposed ColBERT-PRF approach can be exploited in a dense retrieval system, both in end-to-end ranking and reranking scenarios. Dense retrieval approaches, exemplified by ANCE~\cite{xiong2020approximate} and ColBERT~\cite{khattab2020colbert}, are of increasing interest, due to their use of the BERT embedding(s) for representing queries and documents. \craigi{By using directly the BERT embeddings for retrieval, \io{topic drifts} for polysemous words \io{can be} avoided.} To the best of our knowledge, \io{our paper is} the first work investigating PRF in a dense retrieval setting.}

 \section{\mbox{Multi~Representation~Dense~Retrieval}}~\label{sec:DenseRetrieval}

\xiaoi{\io{The queries} and documents \craigi{are} represented \io{by} tokens \craig{from} a vocabulary $V$. \craigi{Each token occurrence has a contextualised} real-valued vector with dimension $d$, called an embedding.}
More formally, let $f: V^{n} \to \mathbb{R}^{n \times d}$ be a function mapping a sequence of terms \nic{$\{t_1, \ldots, t_n\}$, }\xiaoi{representing a query $q$, composed by $|q|$ tokens into a set of embeddings $\{\emb_{q_1}, \ldots, \emb_{q_{|q|}}\}$ and a document composed by $|d|$ tokens into a set of embeddings $\{\emb_{d_1}, \ldots, \emb_{d_{|d|}}\}$}. 
\colbertauths \xiao{~\cite{khattab2020colbert}}
recommended that the number of query embeddings be 32, with extra [MASK]  tokens being used \iadh{as} query augmentation. Indeed, these mask tokens are a differentiable mechanism that allows documents to gain score contributions from embeddings that \craigbeta{do not actually occur} in the query, but which the model assumes could be present in the query.

\looseness -1 \craig{The similarity of two embeddings} is computed by the dot product. Hence, for a query $q$ and a document $d$, their similarity score $s(q,d)$ is obtained by summing the maximum similarity between the query token embeddings and \io{the} document token embeddings~\cite{khattab2020colbert}:
\begin{equation}\label{eq:maxsim}
    s(q,d) = \sum_{i=1}^{|q|}\max_{j=1,\ldots,|d|} \xiaoi{\emb_{q_i}^T \emb_{d_j}}
\end{equation}
\xiaoc{Indeed, Formal et al.~\cite{formal2020white} showed that the dot product $\emb_{q_i}^T \emb_{d_j}$ used by ColBERT implicitly encapsulates token importance, by giving higher scores to tokens that have higher IDF values.}

To obtain a first set of candidate documents, \colbertauths~\cite{khattab2020colbert} make use of FAISS, an approximate nearest neighbour search library, on the pre-computed document embeddings.  Conceptually, FAISS allows to retrieve the $k'$ documents containing the nearest neighbour document embeddings to a query embedding $\xiaoi{\emb_{q_i}}$, \nic{i.e., it provides a function $\mathcal{F}_{d}(\xiaoi{\emb_{q_i}}, k') \xrightarrow{} (d, \dots)$ that returns a list of $k'$ documents, sorted in decreasing approximate \iadh{scores}. }

\looseness -1 However, \craig{these} \nic{approximate} scores are insufficient for accurately depicting the \nic{similarity} scores of the documents, \iadh{hence} the accurate final document scores are computed using Equation~\eqref{eq:maxsim} in a second pass. Typically, for each query embedding, the nearest $k'=1,000$ documents are identified. The union of these documents are reranked using Equation~\eqref{eq:maxsim}. \xiaoi{A separate index data structure (typically in memory) is used to store the \nic{uncompressed} embeddings for each document.} 
%
%
\craigj{To the best of our knowledge, ColBERT~\cite{khattab2020colbert} exemplifies the implementation of an end-to-end IR system that uses multiple representation.} \craigi{Algorithm~\ref{algo:colbert:e2e} summarises the \craigi{ColBERT} retrieval algorithm for the end-to-end dense retrieval \io{approach} proposed by \colbertauths{}, while the top part of Table~\ref{tab:notation} summarises the notation for the main components of the algorithm.}

The easy access to the document embeddings \craigi{used by ColBERT} provides an excellent basis for our dense retrieval \craigbeta{pseudo-relevance feedback} approach. \craigbeta{Indeed, while the use of embeddings (including [MASK] embeddings) in ColBERT  addresses 
the vocabulary mismatch problem, we argue that identifying more related embeddings from \io{the} top-ranked documents may help to further refine the document ranking.} \craig{In particular, as we will show, this permits \xiaoi{representative embeddings} from a set of pseudo-relevant documents to be used to refine the query representation $\phi$.}

\begin{table}[tb]
    \centering
    \caption{Summary of notation -- top group for ColBERT dense retrieval; bottom group for \name{}.}
    \vspace{-.75\baselineskip}
    \begin{tabular}{cp{6.4cm}}
        \toprule
        Symbol & Meaning \\
        \midrule
        $\emb_{q_i}, \emb_{d_j}$ & \xiaoi{An embedding for a query token $q_i$ or a document token~$d_j$}\\
        $\mathcal{F}_{d}(\emb_{q_i}, k')$ & Function returning a list of the $k'$ documents closest to embedding $\emb_{q_i}$\\
        \midrule
        $\Fbembs$ & \xiaoi{Set of feedback embeddings from $\expdoc$ top-ranked feedback documents}\\
        $\centroid_i$ & A \xiaoi{representative (centroid) embedding selected by applying \textsf{KMeans} among $\Fbembs$}\\
        $\clusters$ & Number of representative embeddings to select, i.e., number of clusters for \textsf{KMeans}\\
        $\mathcal{F}_{t}(\centroid_i, r)$ & Function  returning the $r$ token ids corresponding to the $r$ closest document embeddings to embedding~$\centroid_i$\\
        $\IDF_i$ &  \xiaoi{Importance score of $v_i$, calculated as the IDF score of its most likely token }\\
        $F_e$ & \xiaoi{Set of expansion embeddings}\\
        $\expcent$ & Number of \xiaoi{expansion embeddings selected from $\clusters$ representative embeddings} \\
        $\expdoc$ & Number of feedback documents \\ 
        $\fbweight$  & Parameter weighting the contribution of the expansion embeddings\\
        \bottomrule
    \end{tabular}
    \label{tab:notation}\vspace{-\baselineskip}
\end{table}

\begin{algorithm2e}[htb!]
	\DontPrintSemicolon
	\SetKwInOut{Input}{Input}\SetKwInOut{Output}{Output}

	\SetKwProg{myalg}{}{:}{end}
	\SetKwFunction{nameEE}{{\sc ColBERT E2E}}

	\Input{ A query $Q$ }
	\Output{A set $A$ of (docid, score) pairs}
	
	\myalg{\nameEE{{Q}}}
	{
	  \nl $\emb_{q_1}, \ldots, \emb_{q_n} \leftarrow$ {\sf Encode($Q$)} \;
      \nl $D \leftarrow \emptyset$\;
	  \nl \For{$\emb_{q_i}$ {\bf in} $\emb_{q_1}, \ldots, \emb_{q_n}$} {
        \nl $D \leftarrow D \cup \mathcal{F}_{d}(\emb_{q_i}, k')$\;
	  }
      \nl $A \leftarrow \emptyset$\;
	  \nl \For{$d$ {\bf in} $D$} {
	      \nl $s \leftarrow \sum_{i=1}^{|q|}\max_{j=1,\ldots,|d|} \emb_{q_i}^T \emb_{d_j}$\;
	      \nl $A \leftarrow A \cup \big\{(d,s)\big\}$\;
	  }
	  \nl {\bf return} $A$ \;
	}
	\caption{The ColBERT E2E algorithm}
	\label{algo:colbert:e2e}
\end{algorithm2e}



\section{Dense Pseudo-Relevance Feedback}~\label{sec:DenseQE}


\looseness -1 The aim of a pseudo-relevance feedback  approach is typically to generate a refined query representation by \iadh{analysing} the text of the feedback documents. In our \iadh{proposed} \name{} approach, we are inspired by conventional PRF approaches such as \textsf{Bo1}~\cite{amati2002probabilistic} and \textsf{RM3}~\cite{abdulumass},
which assume that good expansion terms will occur frequently in the feedback set \craigi{(and hence \io{are somehow} {\em representative} of the information need underlying the query)}, but \nic{infrequent} in the collection as a whole \craigi{(therefore are sufficiently {\em discriminative})}. Therefore, we aim to encapsulate \craig{these intuitions} while operating in the embedding space $\mathbb{R}^d$, where \iadh{the} exact counting of frequencies is not \iadh{actually} possible. \craigi{\io{In particular, the} bottom part of Table~\ref{tab:notation} summarises the main notations that we use in describing \name{}.}

In \craigbeta{this} section, we detail how we identify  representative (centroid) embeddings from the feedback documents (Section~\ref{ssec:model:clustering}), how we ensure that those centroid embeddings are \craig{sufficiently discriminative} (Section~\ref{ssec:model:idf}), and how we apply these discriminative representative centroid embeddings for (re)ranking (Section~\ref{ssec:model:alg}).  We conclude with an illustrative example (Section~\ref{ssec:model:ex}) and \iadh{a} \craigi{discussion} of the novelty of \name{} (Section~\ref{ssec:model:novelty}).




\subsection{\craig{Representative} Embeddings in Feedback Documents}\label{ssec:model:clustering}

\looseness -1 \iadh{First}, we need to \craig{identify representative} embeddings $\{\centroid_1,\ldots,\centroid_\clusters\}$ among all embeddings in the feedback documents set. \craigi{A typical ``sparse'' PRF approach -- such as RM3 -- would count the frequency of terms occurring in the feedback set to identify representative ones. However, in a dense embedded setting, the document embeddings are not countable. Instead, we resort to clustering to identify patterns in the embedding space that are representative of embeddings.}

\craigi{Specifically, let} \craigbeta{$\Fbembs(q,\expdoc)$ be the set of all document embeddings from the $\expdoc$ top-ranked feedback documents.}
Then, we apply the \textsf{KMeans} clustering algorithm to $\Fbembs(q,\expdoc)$:
\begin{equation}
    \nic{\{\centroid_1, .., \centroid_\clusters \} = \textsf{KMeans}\big(\clusters, \Fbembs(q, \expdoc)\big).}
\end{equation}
\looseness -1 \nic{By applying clustering, we obtain $\clusters$ representative centroid embeddings of the feedback documents.} \craig{Next, we determine how well these centroids discriminate among \io{the} documents in the corpus}.



\subsection{\nic{Identifying Discriminative Embeddings \craig{among} Representative Embeddings}}\label{ssec:model:idf}
Many of the $\clusters$ representative embeddings may represent stopwords and therefore \iadh{are not} \io{sufficiently} informative when retrieving documents. \iadh{Typically}, identifying informative and discriminative expansion terms from feedback documents would involve examining the collection frequency or the document frequency of the constituent terms\xiao{~\cite{roy2019selecting,cao2008selecting}}. However, there may not be a one-to-one relationship between query/centroid embeddings and actual tokens, \xiaow{hence} we seek to map each centroid $\centroid_i$ to a possible token $t$.



\looseness -1 \nic{We resort to FAISS to achieve this, through the function $\mathcal{F}_{t}(\centroid_i, r) \xrightarrow{} (t, \ldots)$ that, given the centroid embedding  $\centroid_i$ and $r$, returns the list of the $r$ \textit{token ids} corresponding to the $r$ closest document embeddings to the centroid.}\footnote{This additional mapping 
can be recorded at indexing time, using the same FAISS index as for dense retrieval, increasing \iadh{the} index \craigi{size} by 3\%.} \craig{From a probabilistic viewpoint, the likelihood $P(t|\centroid_i)$ of a token $t$ given an embedding $\centroid_i$  can be \craigi{obtained} as:}
\begin{equation}\label{eqn:toklookup}
P(t|\xiaoi{\centroid_i}) = \frac{1}{r} \sum_{\tau \in \mathcal{F}_{t}(\centroid_i, r)} \mathbbm{1}[\tau = t],
\end{equation}
\craig{where $\mathbbm{1}[]$ is the indicator function.
In this \craigbeta{work, for simplicity}, we choose the most likely token id, i.e., $t_i = \argmax_t P(t|\centroid_i)$. Mapping back to a token id allows us to make use of \io{Inverse Document Frequency (IDF)}, \iadh{which} can be pre-recorded for each token id. The 
\xiaoi{importance} $\IDF_i$ of \xiaoi{a centroid} 
embedding $\centroid_i$ is obtained using a traditional IDF formula\footnote{\craigi{We \io{have observed} no marked empirical benefits in using other IDF formulations.}}:} 
$\IDF_i = \log \left(\frac{N+1}{N_{i}+1}\right)$,
\looseness -1 \craig{where $N_i$ is the number of passages containing the token $t_i$ and $N$ is the \craigi{total} number of passages in the collection. While this approximation of embedding informativeness is obtained by mapping back to tokens, as we shall show, it is very effective. We leave to future work the derivation of a \iadh{tailored} informativeness measure based upon embeddings alone, for instance using kernel density estimation upon the embedding space.} Finally, we select the $\expcent$ most informative centroids as expansion embeddings  based on the $\IDF_i$ importance scores \xiaoi{as follows:}
\begin{equation}
    F_e = {\sf TopScoring}\Big(\big\{(\centroid_1, \IDF_1),\ldots,(\centroid_K, \IDF_K)\big\},f_e\Big)
\end{equation}
\xiaoi{where {\sf TopScoring($A, c$)} returns the $c$ elements of $A$ with the highest importance score.}

\subsection{Ranking and Reranking with \name{}}\label{ssec:model:alg}
\nic{Given the original $|q|$ query embeddings and the $\expcent$ expansion embeddings, we incorporate the score contributions of the expansion embeddings in Eq.~\eqref{eq:maxsim} as follows:}
\begin{equation}\label{eqn:weightedmaxsum}
    \nic{s(q,d) = \sum_{i=1}^{|q|}\max_{j=1,\ldots,|d|} \xiaoi{\phi_{q_i}^T \phi_{d_j}} + \fbweight \sum_{(\upsilon_i,\sigma_i) \in F_e} \max_{j=1,\ldots,|d|} \IDF_i \xiaoi{\upsilon_{i}^T \phi_{d_j}},}
\end{equation}
\looseness -1 \nic{where $\fbweight > 0$ is a parameter weighting the contribution of the expansion embeddings, and the score produced by each expansion embedding is further weighted \craigi{by the IDF weight of its most likely token, $\IDF_i$}.} \craig{Note that Equation~(\ref{eqn:weightedmaxsum}) can be applied to rerank the documents obtained from the initial query, or as part of a full re-execution of the full dense retrieval operation including the additional }
\xiaoi{$f_e$ expansion embeddings}.
\craig{In both ranking and reranking, \name{} has 4 parameters: $\expdoc$, the number of feedback documents; $\clusters$, the number of clusters; $\expcent \leq \clusters$, the number of expansion embeddings; and $\fbweight$, the importance of the expansion embeddings during scoring.} \xiaoi{\craigi{Furthermore}, we provide the \io{pseudo-code} of our proposed ColBERT PRF ReRanker in Algorithm 2. The \io{ColBERT-PRF} Ranker can be easily obtained by inserting lines 3-4 of \io{Algorithm~1} at line 10 of \io{Algorithm~2}, \craig{and adapting the max-sim scoring in Eq.~\eqref{eq:maxsim} to encapsulate \io{the} original query embeddings as well as the expansion embeddings.}}

\begin{algorithm2e}[htb!]
	\DontPrintSemicolon
	\SetKwInOut{Input}{Input}\SetKwInOut{Output}{Output}

	\SetKwProg{myalg}{}{:}{end}
	\SetKwFunction{nameprf}{{\sc ColBERT PRF}}

	\Input{ A query $Q$,\\number of feedback documents $f_b$,\\number of representative embeddings $K$,\\number of expansion embeddings $f_e$ }
	\Output{A set $B$ of (docid, score) pairs}
	
	\myalg{\nameprf{$Q$}}
	{
	  \nl $A \leftarrow $ {\sf ColBERT E2E($Q)$}\;
      \nl $\Fbembs(Q,\expdoc) \leftarrow $ set of all document embeddings from\\\hspace{1.25cm} the $\expdoc$ top-scored documents in $A$\;
      \nl $V \leftarrow \emptyset$\;
	  \nl $\upsilon_1, \dots, \upsilon_K$ = {\sf KMeans\big($K$, $\Phi(Q,f_b)$\big)} \;

	  \nl \For{$\upsilon_i$ {\bf in} $\upsilon_1, \dots, \upsilon_K$} {
	    \nl $t_i \leftarrow \argmax_t \frac{1}{r} \sum_{\tau \in \mathcal{F}_{t}(\centroid_i, r)} \mathbbm{1}[\tau = t]$ \;  
	    \nl $\sigma_i \leftarrow  \log \left(\frac{N+1}{N_{i}+1}\right)$\;
	    \nl $V \leftarrow V \cup \big\{(\upsilon_i,\sigma_i)\big\}$\;
	  }
	  \nl $F_e \leftarrow ${\sf TopScoring($V,f_e$)} \;

      \nl $B \leftarrow \emptyset$\;	  
	  \nl \For{$(d, s)$ {\bf in} $A$} {
	      \nl $s \leftarrow s + \fbweight \sum_{(\upsilon_i,\sigma_i) \in F_e}\max_{j=1,\ldots,|d|} \IDF_i \upsilon_i^T \emb_{d_j}$ \;
	      \nl $B \leftarrow B \cup \big\{(d,s)\big\}$\;
	  }
      \nl {\bf return} $B$ \;
	}
	\caption{The ColBERT PRF \craigi{(reranking)} algorithm}\label{algo:colbert:prf}
\end{algorithm2e}


\subsection{Illustrative Example}\label{ssec:model:ex}

\begin{figure}
\centering
\def\scalefactor{0.32}
\def\figsize{45mm}
\begin{subfigure}[b]{42mm}\centering
\includegraphics[width=\figsize]{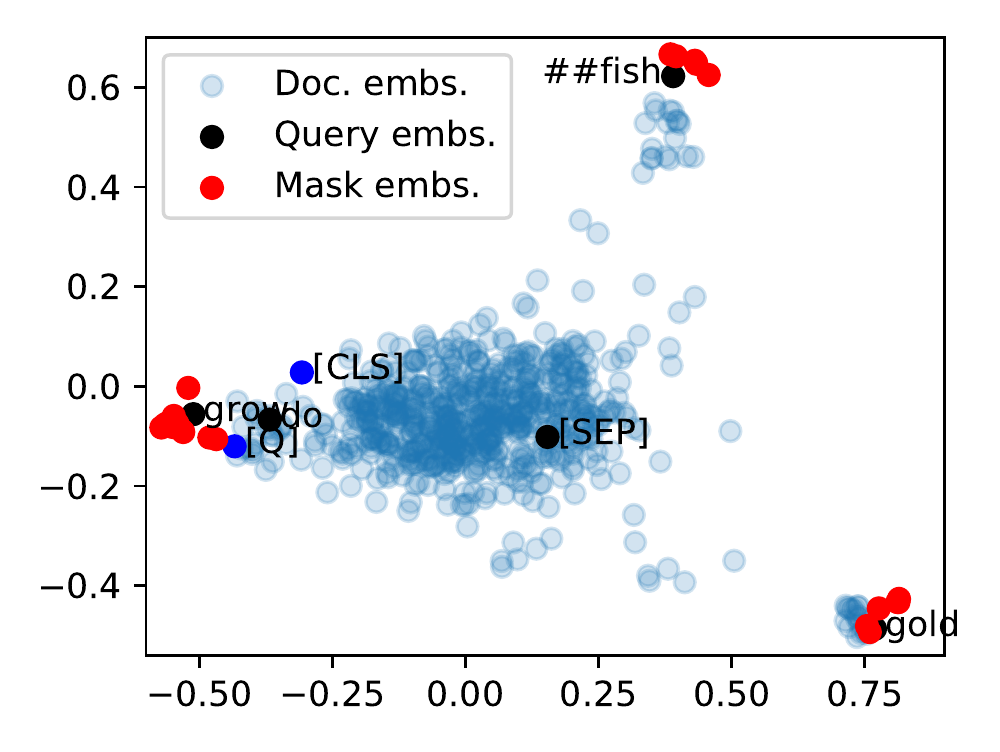}\vspace{-.8\baselineskip}
\caption{Query \& doc. embeddings.}\label{fig:pca:query}
\end{subfigure}
\begin{subfigure}[b]{42mm}\centering
\includegraphics[width=\figsize]{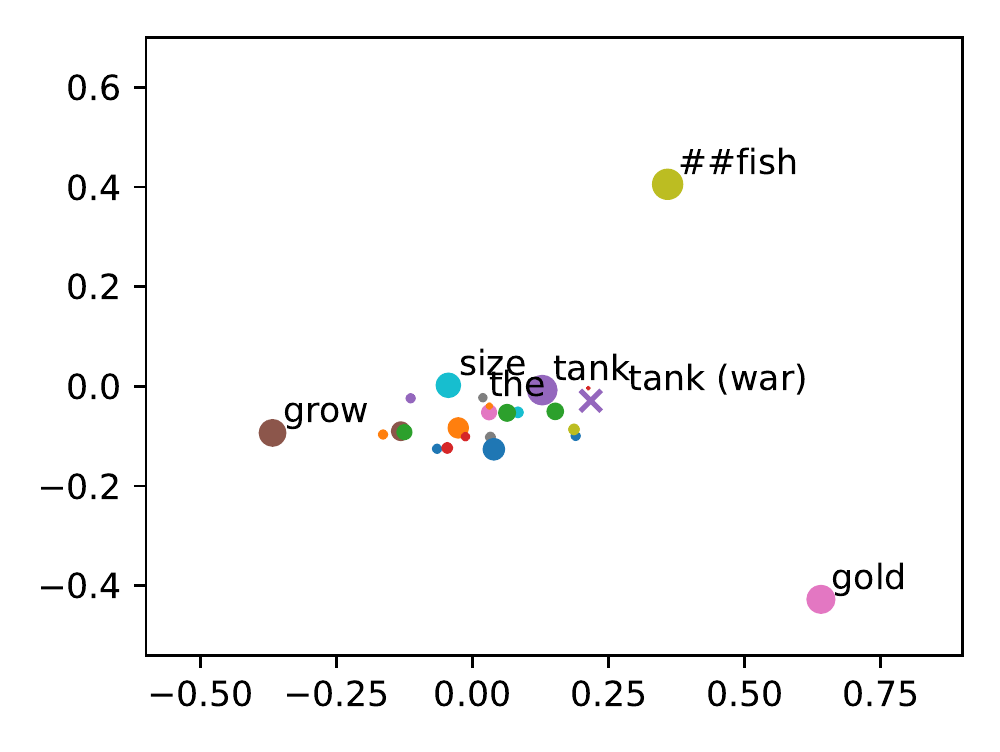}\vspace{-.8\baselineskip}
\caption{Cluster centroids, $\clusters=24$.}\label{fig:pca:doc}
\end{subfigure}
\vspace{-1.2\baselineskip}\caption{\looseness -1 Example showing how \name{} operates for the query `do goldfish grow' in a 2D PCA space. In Figure~\ref{fig:pca}(b), \iadh{the} point size is representative of IDF; five high IDF and one low IDF centroids are shown. \craigj{For contrast, $\times$ `tank (war)' denotes the embedding of `tank' occurring in a non-fish context.}
}\label{fig:pca}\vspace{-\baselineskip}
\end{figure}

We now illustrate the effect of \name{} upon one query from the TREC 2019 Deep Learning track, `do goldfish grow'. We use PCA to quantize the 128-dimension embeddings into 2 dimensions \craigi{purely to allow visualisation}. Firstly, Figure~\ref{fig:pca}(a)  shows the embeddings of the original query \craigbeta{(black ellipses)}; the red [MASK] tokens \craigbeta{are also visible,} clustered around the original query terms (\#\#fish, gold, grow). \xiaow{Meanwhile, document embeddings extracted from 10 feedback documents are shown as light blue \craigbeta{ellipses} in Figure~\ref{fig:pca}(a). } 
There appear to be visible \xiao{clusters} \craigbeta{of document embeddings} near
the query embeddings, but also other document embeddings exhibit some clustering. The mass of embeddings near the origin \xiaow{is} not \xiao{distinguishable} in PCA. \craig{Figure~\ref{fig:pca}(b) demonstrates the application of KMeans clustering upon the document embeddings; we map back to the original tokens by virtue of Equation~\eqref{eqn:toklookup}. In Figure~\ref{fig:pca}(b), \iadh{the} point size is indicative of the IDF of the corresponding token. We can see \xiao{that the cluster centroids} with high IDF correspond to the original query tokens (`gold', `\#\#fish', `grow'), as well as the related terms (`tank', `size'). In contrast, a centroid with low IDF is `the'. This \io{illustrates} the utility of our proposed \name{} approach in using KMeans to identify representative clusters of embeddings, as well as using IDF to differentiate useful clusters.}

\looseness -1 \craigj{Furthermore, Figure~\ref{fig:pca}(b) also includes, marked by an $\times$ and denoted `tank (war)', the embedding for the word \yaya{`tank'} when placed in the passage {\em``While the soldiers advanced, the tank bombarded the troops with artillery''}. It can be seen that, even in the highly compressed PCA space, the `tank' centroid embedding is distinct from the embedding of `tank (war)'. This shows the utility of \name{} \yaya{when} operating in the embedding space, as the PRF process for the query `do goldfish grow' will not retrieve documents containing `tank (war)', but will focus on a fish-related context, thereby dealing with the polysemous nature of a word such as `tank'. To the best of our knowledge, this is a unique feature of \name{} among PRF approaches.}



\subsection{Discussion}\label{ssec:model:novelty}
\craig{To the best of our knowledge \name{} is the first investigation of pseudo-relevance feedback for dense retrieval. Existing works on \craigi{neural} pseudo-relevance feedback, such as Neural PRF~\cite{li2018nprf} and BERT-QE~\cite{zheng2020bert} only function as rerankers. Other approaches such as DeepCT~\cite{dai2020context} and doc2query~\cite{nogueira2019document,nogueira2019doc2query} use neural models to augment documents before indexing using a traditional inverted index. \craig{CEQE~\cite{naseri2021ceqe} generates words to expand the initial query, which is then executed on the inverted index}. \craigj{However, returning the BERT embeddings back to textual word forms can result in polysemous words negatively affecting retrieval.} In contrast, \name{} operates entirely on an existing dense index representation (without augmenting documents), and can function for \iadh{both} ranking as well as reranking.} \craigj{By retrieving using feedback embeddings directly, \name{} addresses polysemous words (such as `tank', illustrated above). Furthermore,} \craigbeta{it also requires no  additional neural network training beyond that of ColBERT.} 


\section{Experimental Setup}~\label{sec:expsetup}
\noindent Our experiments address the \io{three} following research questions:\\

\xiaoi{$\bullet$ RQ1: Can a multiple representation dense retrieval approach be enhanced by \io{pseudo-relevance} feedback, i.e. \io{can} ColBERT-PRF \io{outperform} ColBERT dense retrieval?}

$\bullet$ RQ2: How does ColBERT-PRF \io{compare} to other existing baseline and state-of-the-art approaches, namely:

\noindent (a) lexical (sparse) baselines, including using PRF,

\noindent (b) neural augmentation approaches, namely DeepCT and docT5query,


\noindent (c) \xiaoi{BERT-QE Reranking models?}


$\bullet$ RQ3: \craigbeta{What is the impact of the parameters of \name{}, namely \craigd{the} number of clusters and expansion embeddings, the number of feedback documents and the $\fbweight$ parameter controlling the influence of the expansion embeddings?}






\subsection{Dataset \& Measures}
\looseness -1 Experiments are conducted on the MSMARCO passage corpus, using the TREC 2019 Deep Learning track topics (43 topics with \io{an} average \io{of} \xiao{95.4} relevance judgements per query), \iadh{as well as} \xiao{the TREC 2020 Deep Learning track topics (54 topics with \io{an} average \io{of} 66.8 relevance judgements per query).} \craig{We omit \iadh{topics from} the MSMARCO Dev set, which have only sparse judgements, $\sim$1.1 per query. \iadh{Indeed, pseudo}-\iadh{relevance} feedback approaches are known \iadh{to be not} effective on test collections with few judged documents~\cite{amati2004query}}.

\looseness -1 We report the commonly used metrics for \iadh{the} \craig{TREC 2019 and TREC 2020 query sets following the corresponding \iadh{track overview} papers~\cite{craswell2020overview,craswell2021overview}:} we report mean reciprocal rank (MRR) and normalised discounted cumulative gain (NDCG) calculated at rank 10, as well as \craigc{Recall and} Mean Average Precision (MAP) at rank 1000\craigc{~\cite{craswell2020overview}}. \craig{For \iadh{the} \craigc{MRR, MAP and Recall} metrics, we treat passages with label grade 1 as non-relevant, following~\cite{craswell2020overview,craswell2021overview}.} \xiaoi{In addition, we also report the Mean Response Time (MRT) for each retrieval system.}
\xiao{For significance testing, we use the paired t-test ($p<0.05$) and apply \iadh{the} Holm-Bonferroni multiple testing correction.}


\subsection{Implementation and Settings}
\looseness -1 We conduct experiments using \craigc{PyTerrier~\cite{pyterrier} and, in particular using our PyTerrier\_ColBERT plugin\footnote{\url{https://github.com/terrierteam/pyterrier_colbert}}, which includes \name{} as well as our adaptations of the ColBERT source code.}
In terms of \iadh{the} ColBERT configuration, we train ColBERT upon the MSMARCO passage ranking triples file for 44,000 batches, \craigbeta{applying the} parameters specified by \colbertauths{} in~\cite{khattab2020colbert}:
\craigi{Maximum document length is set to 180 tokens and queries are encoded into 32 query embeddings (including [MASK] tokens); We encode all passages to a FAISS index that has been trained using 5\% of all embeddings;} At retrieval time, FAISS retrieves $k'=1000$ document embeddings for every query embedding.

\craig{\name{} is implemented using the KMeans implementation~\cite{kmeansPlusPlus} of sci-kit learn (sklearn).} For query expansion settings, we follow the default settings of Terrier~\cite{ounis2005terrier}, which is 10 expansion terms obtained from 3 feedback documents; we follow the same default setting for \name{}, additionally using \craigbeta{representative values, namely} \xiao{$\clusters=24$} clusters\footnote{\craigj{Indeed, $K=24$ gave reasonable looking clusters in our initial investigations, and, as we shall see in Section 6.3, is an effective setting for the TREC 2019 query set.}}, \craig{and \craigbeta{$\fbweight=\{0.5, 1\}$} for the weight of the expansion embeddings}. We later show the impact of these parameters \io{when we address}
\xiaoi{RQ3}.



\subsection{Baselines}\label{ssec:expsetup:baselines}
\looseness -1 \craigbeta{To} test the effectiveness of our proposed dense \xiao{PRF} \craigi{approach}, we compare with \xiaoi{four families of baseline models, for which we vary the \craigi{use of a BERT-based reranker (namely \craigc{BERT or} ColBERT)}. \craigc{For the BERT reranker, we use OpenNIR~\cite{macavaney:wsdm2020-onir} and {\tt capreolus/ bert-base-msmarco} fine-tuned model from~\cite{li2021parade}.}} For the ColBERT reranker, \craigc{unless otherwise noted, we use the existing pre-indexed ColBERT representation of documents for efficient reranking. The four families are:} 

\noindent{\em Lexical Retrieval Approaches:} \xiaoi{These are traditional retrieval models using a sparse inverted index, with and without \craigc{BERT and ColBERT \io{rerankers}, namely: (i) BM25 (ii) BM25+BERT (iii) BM25+ColBERT, (iv) BM25+RM3, (\xiaoc{v}) BM25+RM3+\-BERT and (\xiaoc{vi}) BM25+RM3+\-ColBERT.}}


\looseness -1 \noindent{\em Neural Augmentation Approaches:} These use neural components to augment the (sparse) inverted index: (i) BM25+DeepCT and (ii)  BM25+docT5query, \craigc{both without and with BERT and ColBERT rerankers}. \craigc{For BM25+docT5\-query+ColBERT, the ColBERT reranker is applied on \craigc{expanded} document texts encoded at querying time, rather than \io{the} indexed ColBERT representation. The response time for BM25+docT5query+ColBERT reflects this difference.}

    
\noindent\textit{Dense Retrieval Models:} \xiaoi{This family consists of the dense retrieval approaches:} (i) ANCE: The ANCE~\cite{xiong2020approximate} model is a single representation dense retrieval model. \craig{We use the trained models provided by the authors trained on MSMARCO training data.} \iadh{(ii)} ColBERT E2E: ColBERT end-to-end (E2E)~\cite{khattab2020colbert} is the dense retrieval version of ColBERT, as defined in Section~\ref{sec:DenseRetrieval}.


\noindent\textit{BERT-QE Models:} 
We apply BERT-QE~\cite{zheng2020bert} on top of a strong sparse baseline and our dense retrieval baseline, ColBERT E2E, i.e. (i) {BM25+RM3+ColBERT+BERT-QE} and (ii) {ColBERT E2E+BERT-QE}; Where possible, we use the ColBERT index for scoring passages; for identifying the top scoring chunks within passages, we use ColBERT in a slower ``text'' mode, \craigi{i.e.\ without using the index}. For \io{the} BERT-QE parameters, we follow the settings in \cite{zheng2020bert}, in particular using the recommended settings of $\alpha=0.4$ and $\beta=0.9$, which 
\xiaoj{are} also the most effective on MSMARCO. Indeed, to the best our knowledge, this is the first application of BERT-QE upon dense retrieval, \io{the} first application of BERT-QE on MSMARCO and \io{the} first application using ColBERT. \craigc{We did attempt to apply BERT-QE using the BERT re-ranker, but we found it to be ineffective on MSMARCO, and exhibiting a response time exceeding 30 seconds per query, hence we omit it from our experiments.}

Note that, \craigc{at the time of writing}, there is no publicly available implementation of the very recent CEQE~\cite{naseri2021ceqe} approach, \craigc{and hence} we omit it from our experiments. \craigc{Code to reproduce \name{} is available in the PyTerrier\_ColBERT repository.\footnotemark[3]}
 

\begin{table*}[tb]
    \centering
    \caption{Comparison with baselines. Superscripts \xiaoc{a...p}
    denote significant improvements over the indicated baseline model(s). The highest value in each column is boldfaced. \craigi{The higher MRT of BM25+docT5query+ColBERT is expected, as we do not have a ColBERT index for the docT5query representation.} }
    \vspace{-.75\baselineskip}
    \begin{adjustbox}{max width=178mm}
\begin{tabular}{lcccccccccc}
 \toprule 
 &  \multicolumn{5}{c}{TREC 2019 (43 queries)} & \multicolumn{5}{c}{TREC 2020 (54 queries)} \\
 \cmidrule(lr){2-6}\cmidrule(lr){7-11}
 {}   &  MAP  &NDCG@10 &MRR@10  & \xiaoc{Recall} & MRT &MAP  &NDCG@10 &MRR@10  & \xiaoc{Recall} & MRT
 \\
\midrule
\multicolumn{11}{c}{Lexical Retrieval Approaches}\\
\midrule
 
 BM25 (a)
 & 0.2864 & 0.4795 & 0.6416  &0.7553 & \nic{132.4} 
 & 0.2930 & 0.4936 & 0.5912  &0.8103 & \nic{129.1} 
 \\
   \xiaoc{BM25+BERT} (b)
   &0.4441 &0.6855 &0.8295 &0.7553 &3588.98
   &0.4699 &0.6716 &0.8069 &0.8103 &3553.36
  
  \\
  
   BM25+ColBERT (c) 
  &0.4582  &0.6950  &0.8580  &0.7553 & \nic{201.9} 
  &0.4752  &0.6931  &0.8546  &0.8103 & \nic{203.2} 
  \\

 BM25+RM3 (d) 
 &0.3108 &0.5156 & 0.6093  &0.7756 & \nic{201.4}
 &0.3203 &0.5043 & 0.5912  &0.8423 & \nic{247.7} 
 \\
  \xiaoc{BM25+RM3+BERT} (e)
 & 0.4531 &0.6862 &0.8275 &0.7756 &4035.01
 & 0.4739 &0.6704 &0.8079 &0.8423 &4003.09
  \\
 
   BM25+RM3+ColBERT (f) 
  &0.4709  &0.7055   &0.8651   &0.7756  & \nic{319.9}
  &0.4800  &0.6877   &\bf{0.8560}   &0.8423 & \nic{228.2} 
  \\

\midrule
\multicolumn{11}{c}{Neural Augmentation Approaches}\\
\midrule
 
 BM25+DeepCT (g) 
 & 0.3169 &0.5599 &0.7155  &0.7321 & \nic{54.3} 
 & 0.3570 &0.5603 &0.7090  &0.8008 & \nic{63.8} 
 \\
  \xiaoc{BM25+DeepCT+BERT}~(h) 
 &0.4308 &0.7011 &0.8483 &0.7321 &3736.35
 &0.4671 &0.6852 &0.8068 &0.8008 &3718.29
  
  \\
   BM25+DeepCT+ColBERT~(i)
  & 0.4416 &0.7004 & 0.8541  &0.7321 & \nic{129.4} 
  & 0.4757 &0.7071 & 0.8549  &0.8008& \nic{140.5} 
  \\

 BM25+docT5query~(j)
 & 0.4044 & 0.6308 &0.7614  &0.8263 & \nic{281.8} 
 & 0.4082 & 0.6228 &0.7434 &0.8456 & \nic{295.4} 
 \\
 
  \xiaoc{BM25+docT5query+BERT}~(k)
& 0.4802 &0.7123 &0.8483 &0.8263 &8024.90
& 0.4714 &0.6810 &0.8160 &0.8456 &3887.78

  \\
  
   BM25+docT5query+ColBERT~(l)
  &0.5009 &0.7136 & 0.8367  &0.8263 & 2361.88 
  &0.4733 &0.6934 & 0.8021  &0.8456 & 2381.09  
  \\


\midrule
\multicolumn{11}{c}{Dense Retrieval Models}\\
\midrule
  ANCE~(m) 
&0.3715 & 0.6537    & 0.8590 &0.7571    & \nic{199.2}
&0.4070 & 0.6447    & 0.7898 &0.7737    & \nic{178.4}
  \\
  
  ColBERT E2E~(n) 
  & 0.4318 &0.6934 &0.8529    &0.7892   & \nic{599.6} 
  &0.4654  &0.6871  &0.8525   &0.8245   & \nic{530.8} 
  \\
  
\midrule
\multicolumn{11}{c}{BERT-QE Reranking Models}\\
\midrule





  BM25 + RM3 + ColBERT +  BERT-QE~(o) 
   & \xiaoc{0.4832} & 0.7179 &   0.8754  & 0.7756  & 1129.88
    & \xiaoc{0.4842} & 0.6909 &  0.8315  & 0.8423  & 1595.12
  \\
  ColBERT E2E + BERT-QE~(p) 
  & \xiaoc{0.4423} & 0.7013 & 0.8683  &0.7892 &  1260.92    
  & \xiaoc{0.4749} & 0.6911 &  0.8315  &0.8245  & 1327.88 
  
  \\
\midrule
\multicolumn{11}{c}{\name{} Models}\\
\midrule
  \name{} Ranker ($\fbweight$=1)
&\bf{0.5431$^{abcdghijmnp}$} & 0.7352$^{adg}$ &0.8858$^{ad}$   &\xiaoc{0.8706$^{abhmn}$}  &\craigc{4391.21} 
&0.4962$^{adgjm}$  &0.6993$^{adg}$ &0.8396$^{ad}$    &\xiaoc{\bf{0.8892}$^{abghlmn}$} &\craigc{4233.34} 
  \\
  
  \name{} ReRanker ($\fbweight$=1)
&0.5040$^{adgmnp}$ & 0.7369$^{adg}$ & 0.8858$^{ad}$     &\xiaoc{0.7961}  & \craigc{3598.23}
&0.4919$^{adgj}$ &0.7006$^{adg}$ & 0.8396$^{ad}$  &\xiaoc{0.8431$^{m}$} & \craigc{3607.18}
  \\

   \name{} Ranker ($\fbweight$=0.5)
  &0.5427$^{abcdghijmnp}$  & 0.7395$^{adgjm}$   &\bf{0.8897}$^{ad}$   &\xiaoc{\bf{0.8711}$^{abhmn}$} & \craigc{4132.30}
  &\bf{0.5116$^{adgjmn}$}  & 0.7153$^{adgj}$   &{0.8439}$^{ad}$    &\xiaoc{0.8837$^{aghlmn}$}  & \craigc{4300.58}

  \\
  
  \name{} ReRanker ($\fbweight$=0.5)
&0.5026$^{adgmnp}$ & \bf{0.7409}$^{adgjm}$& \bf{0.8897}$^{ad}$  &\xiaoc{0.7977} & \craigc{3576.62}
&0.5063$^{adgjm}$   &\bf{0.7161}$^{adgj}$      &{0.8439}$^{ad}$   &\xiaoc{0.8443$^{m}$ }  & \craigc{3535.69} 
\\

 \bottomrule
\end{tabular}
\end{adjustbox}
\label{tab:RQ1}\vspace{-0.5\baselineskip}
\end{table*}


\section{Results}\label{sec:results}

In the following, we analyse the performance of \name{} wrt. RQs 1-3. For RQs 1 \& 2 (Sections~\ref{ssec:res:rq1} \& \ref{ssec:res:rq2}), we analyse  Table~\ref{tab:RQ1}, which \craigbeta{reports} the \iadh{performances} of all the baselines \io{as well as the} \name{} models on the 43 TREC 2019 \& 54 TREC 2020 queries. Baselines are grouped by the retrieval families discussed in Section~\ref{ssec:expsetup:baselines}.


\subsection{Results for RQ1}\label{ssec:res:rq1}

\looseness -1 \xiaoi{In this section, we examine} the effectiveness of \craigi{a} \io{pseudo-relevance} feedback technique for \craig{the ColBERT} dense retrieval model. On analysing Table~\ref{tab:RQ1}, we first note that \io{the} ColBERT dense retrieval \io{approach} outperforms the single representation dense retrieval model, i.e.\ ANCE for all metrics on both test query sets, probably because the single-representation used in ANCE provides limited information for matching queries and documents~\cite{luan2020sparse}. Based on this, we then compare the \io{performances} of our proposed \name{} models, instantiated as \name{} Ranker
\& \name{} ReRanker, with the more effective ColBERT E2E model. We find that both \io{the} Ranker and ReRanker models outperform ColBERT E2E on all the metrics for both \io{used} query sets. Typically, on \craigi{the} TREC 2019 test queries, both \craigi{the} Ranker and ReRanker models exhibit \io{significant improvements} in terms of MAP \craigi{over} the ColBERT E2E model. In particular, we observe a \xiaow{26}\% increase in MAP on TREC 2019\footnote{\craigbeta{Indeed, this is 8\% higher than the highest MAP among all TREC 2019 participants~\cite{craswell2020overview}}.} and \xiaoi{10\%} for TREC 2020 over ColBERT E2E for \io{the} \name{} Ranker. In addition, both \name{} Ranker and ReRanker exhibit \io{significant improvements} over ColBERT E2E in terms of NDCG@10 on TREC 2019 queries. 

The high effectiveness of \name{} Ranker \craigj{(which is indeed higher than \name{} ReRanker) can} \iadh{be explained in that the} expanded query obtained using the PRF process introduces more relevant documents, thus \iadh{it} increases recall after re-executing the query on the dense index. \xiaoc{\craigc{As can be seen from} Table~\ref{tab:RQ1}, ColBERT-PRF Ranker exhibits significant \craigc{improvements} over both ANCE and ColBERT E2E models on Recall.} On the other hand, the effectiveness of \name{} ReRanker also \iadh{suggests} that the expanded query provides a better query representation, \iadh{which} can better rank documents in the existing candidate set. 
\xiaoi{
We further investigate the actual expansion tokens using \io{the} \name{} model.
Table~\ref{tab:Q_exp} lists three example queries from both \io{the} TREC 2019 and 2020 query sets and their tokenised forms as well as the expansion tokens generated by \io{the} \name{} model. For a given query, we used our default setting \io{for the} \name{} model, i.e. \craigj{selecting} ten expansion embeddings; \craigi{Equation~\eqref{eqn:toklookup} is used to resolve embeddings to tokens.} We find that most of the expansion tokens identified are credible supplementary information for each user query and \craigj{can indeed clarify the information needs}.} 
\xiao{Overall, in response to \xiaoi{RQ1}, we conclude that our proposed \name{} model is effective \iadh{compared} to the ColBERT \craigi{E2E} dense retrieval model.}




\subsection{Results for RQ2}\label{ssec:res:rq2}

\looseness -1 Next, to address RQ2(a)-(c), we analyse the \io{performances} of \iadh{the} \name{} Ranker and \name{} ReRanker approaches \io{in comparison} to different groups of baselines, namely sparse (lexical) retrieval approaches, neural augmented baselines, and BERT-QE.

\looseness -1 For RQ2(a), we compare the \name{} Ranker and ReRanker models with the lexical retrieval approaches. For both query sets, both Ranker and ReRanker provide significant \io{improvements} on all evaluation measures compared to the BM25 and BM25+RM3 models.  This is mainly due to the more effective contexualised representation employed in \io{the} \name{} models than the traditional sparse representation used in \io{the} lexical retrieval approaches. \io{Furthermore}, both \name{} Ranker and ReRanker outperform the sparse retrieval approaches when reranked by \xiaoc{\craigc{either the BERT or}} the ColBERT \craigc{models} -- 
\xiaoc{e.g.} \craigc{BM25+(Col)BERT and BM25+RM3+(Col)BERT} -- on all metrics. In particular, \name{} Ranker exhibits marked improvements over the BM25 with \xiaoc{BERT or }ColBERT  reranking approach for MAP on \io{the} TREC 2019 queries. \craigc{This indicates that our query expansion in the contextualised embedding space produces query representations that result in improved retrieval effectiveness.}
Hence, in answer to RQ2(a), we find that our proposed \name{} models show significant \io{improvements} \io{in} retrieval effectiveness over sparse baselines.

\looseness -1 For RQ2(b), on analysing the neural augmentation approaches, we observe \io{that} both \io{the} DeepCT and docT5query neural components could lead \io{to} effectiveness \io{improvements} over the corresponding lexical retrieval 
\xiaoj{models} without neural augmentation. However, despite their improved effectiveness, our proposed \name{} models exhibit marked improvements over the neural augmentation approaches. \craigi{Specifically}, on the  TREC 2019 query set, \xiaoi{\name{} Ranker significantly outperforms} \xiaoc{4 out of 6 neural augmentation baselines and the BM25+DeepCT baseline on MAP.} 
Meanwhile, both \name{} Ranker and ReRanker exhibit significant improvements over \craigc{BM25+DeepCT and BM25+docT5query}
on MAP for TREC 2020 queries, \craigc{and exhibit improvements upto 9.5\% improvements over neural augmentation approaches with neural re-ranking (e.g. MAP 0.4671$\rightarrow$0.5116).} On analysing these comparisons, the effectiveness of \io{the} \name{} models \io{indicates} that the query representation enrichment in a contextualised embedding space leads \io{to a} higher effectiveness performance than the sparse representation document enrichment. Thus, in response to RQ2(b), \io{the} \name{} models exhibit markedly higher \io{performances} than the neural augmentation approaches.

We further compare \io{the} \name{} models with the recently proposed BERT-QE Reranking model. In particular, we provide results when using BERT-QE to rerank both BM25+RM3 as well as ColBERT E2E. Before comparing \io{the} \name{} models with \io{the} BERT-QE rerankers, we first note that BERT-QE doesn't provide benefit to MAP on either query set, but can lead \io{to} a marginal improvement for NDCG@10 and MRR@10. However, the BERT-QE reranker models still \io{underperform} compared to our  \name{} models. Indeed, ColBERT E2E+BERT-QE exhibits \io{a} performance significantly lower than both \name{} Ranker and ReRanker on the TREC 2019 query set. Hence, in response to RQ2(c), we find that \io{the} \name{} models significantly outperform \io{the} BERT-QE reranking models.

\craigi{Finally, we consider the mean response times reported in Table~\ref{tab:RQ1}, noting that ColBERT PRF exhibits higher response times than other \craigc{ColBERT-based baselines, and similar to BERT-based re-rankers}. There are several reasons for \craigc{ColBERT PRF's speed}: Firstly, the KMeans clustering of the feedback embeddings is conducted online, and the scikit-learn implementation we used is fairly slow -- we tried other markedly faster KMeans implementations, but \io{they} were limited in terms of effectiveness (particularly for MAP), perhaps due to the lack of the KMeans++ initialisation procedure~\cite{kmeansPlusPlus}, which scikit-learn adopts; Secondly ColBERT PRF adds more expansion embeddings to the query - for the ranking setup, each feedback embedding can potentially cause a further $k'=1000$ documents 
to be scored - \craigc{further tuning of ColBERT's $k'$ parameter may allow efficiency improvements for \name{} without much loss of effectiveness, at least for the first retrieval stage}. Overall, we contend that the effectiveness benefits exhibited by \name{} demonstrate the promise of our approach, and hence we leave further studies of \craigj{improving the efficiency of \name{} to future work, such as through \yaya{deploying} efficient clustering algorithms (either online, i.e., per-query, or offline, i.e., based on the existing FAISS index), as well as more conservative second-pass retrieval settings}}.

\begin{figure}
\centering
\def\scalefactor{0.32}
\def\figsize{45mm}
\begin{subfigure}[b]{42mm}\centering
\includegraphics[width=\figsize]{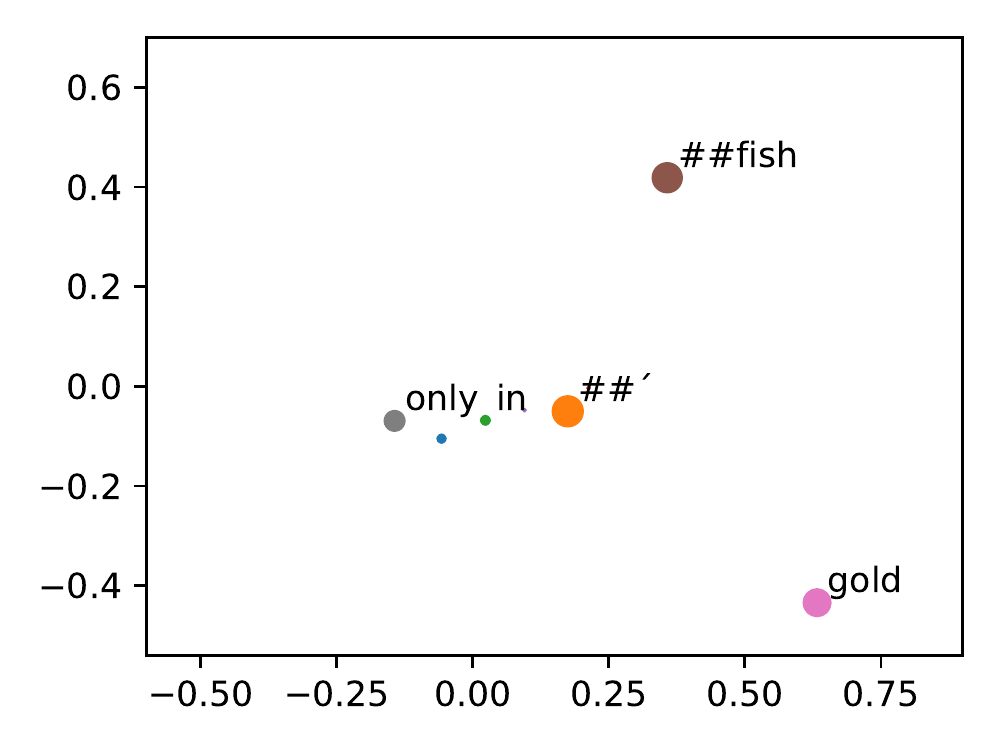}\vspace{-.8\baselineskip}
\caption{Cluster centroids, $\clusters=8$.}\label{fig:pca:doc3}
\end{subfigure}
\begin{subfigure}[b]{42mm}\centering
\includegraphics[width=\figsize]{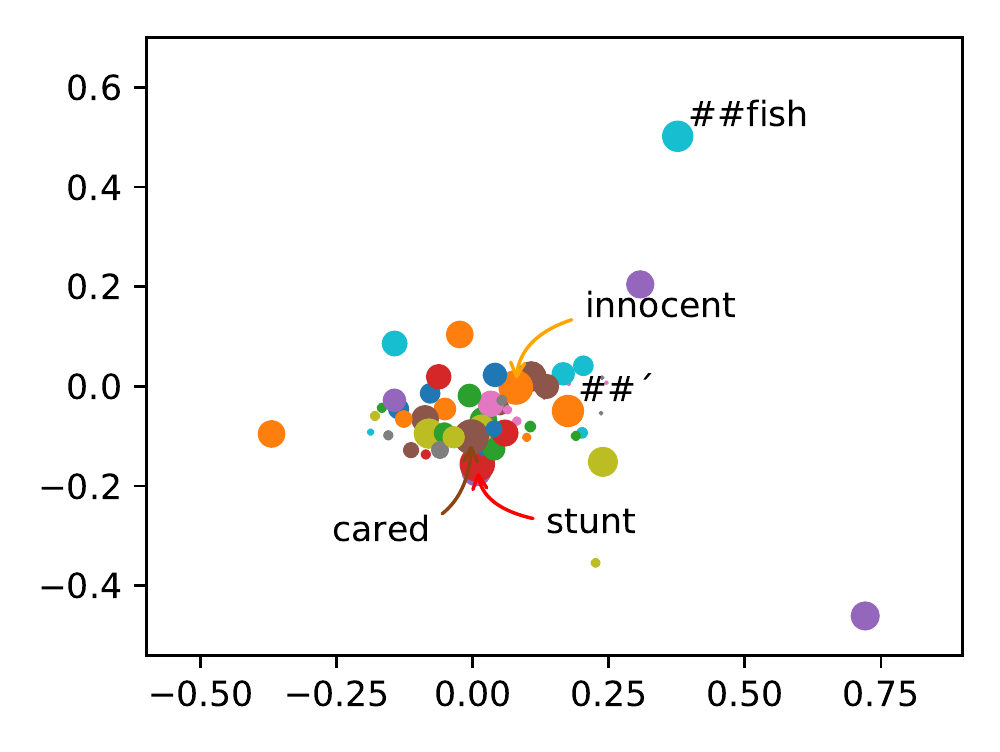}\vspace{-.8\baselineskip}
\caption{Cluster centroids, $\clusters=64$.}\label{fig:pca:doc4}
\end{subfigure}
\vspace{-1.2\baselineskip}\caption{\xiaoi{Embeddings selected using different number of clustering centroids $\clusters$ for the query `do goldfish grow', the point size is representative of IDF.}
}\label{fig:expK}\vspace{-\baselineskip}
\end{figure}


\begin{figure*}[t]
\centering
\def\scalefactor{0.24}
\def\figsize{48mm}
\begin{subfigure}[b]{\scalefactor\linewidth}\centering
\includegraphics[width=\figsize]{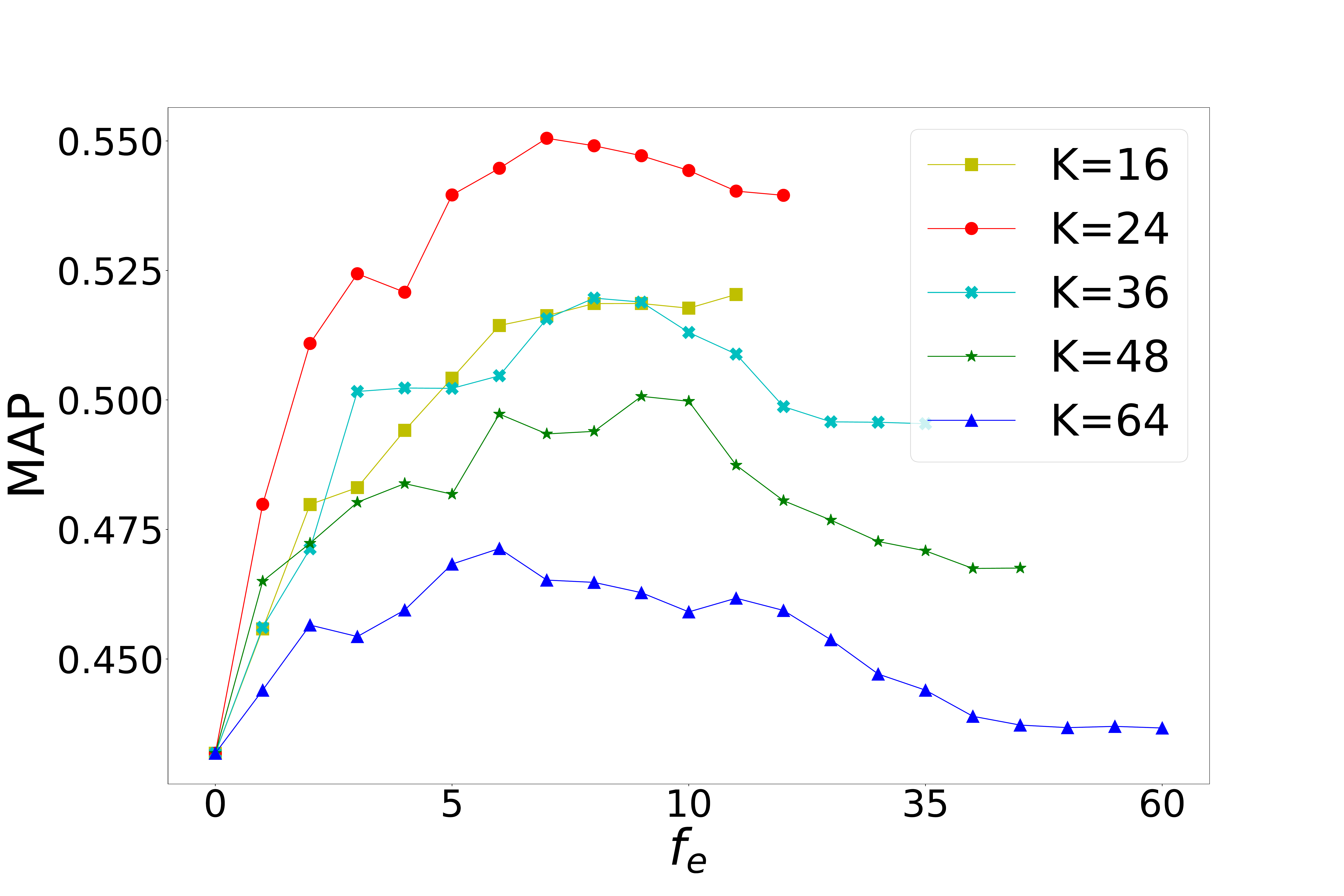}
\caption{Impact of $\clusters$ and $\expcent$ on \name{} Ranker.}\label{fig:rankerexp}
\end{subfigure}
\begin{subfigure}[b]{\scalefactor\linewidth}\centering
\includegraphics[width=\figsize]{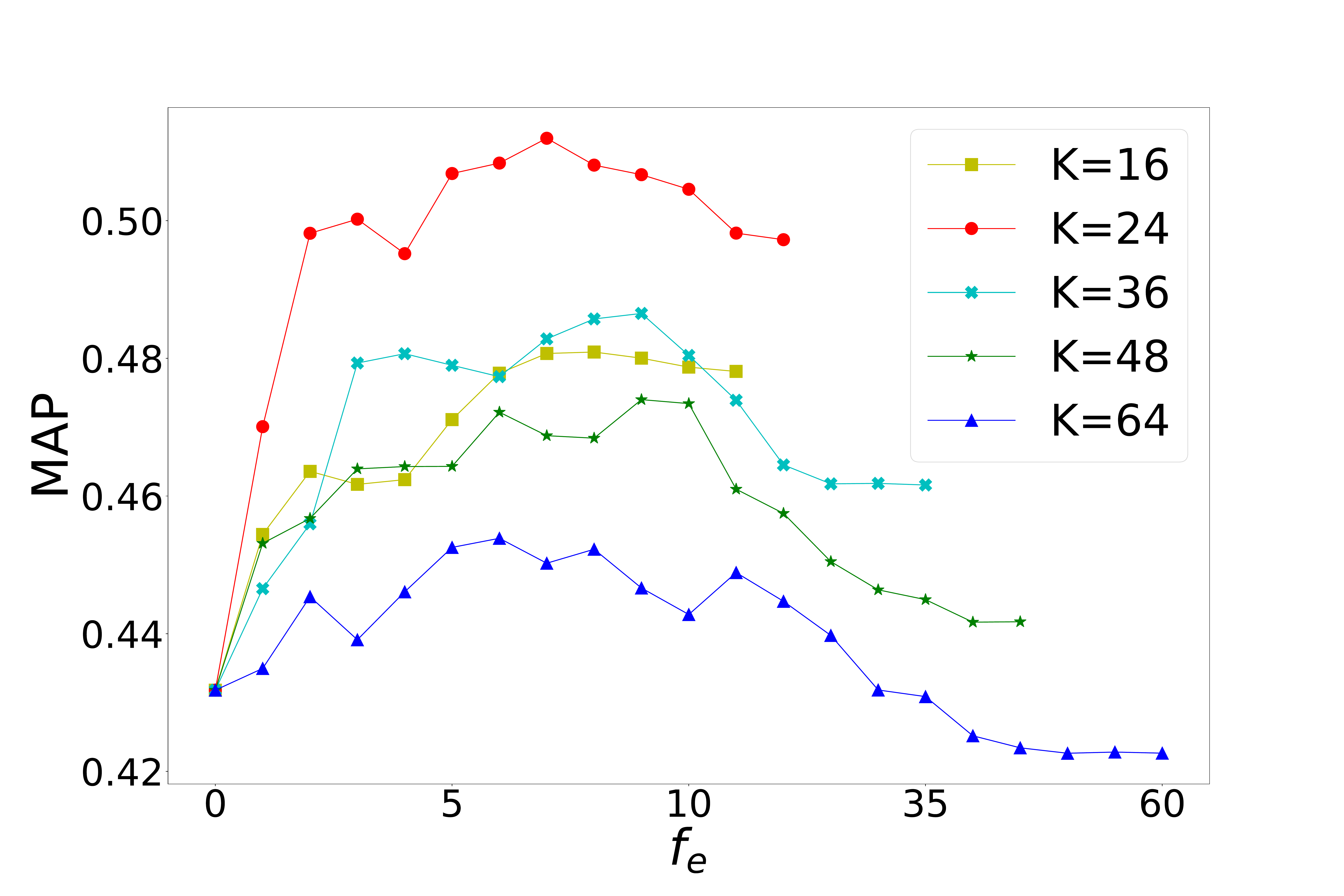}
\caption{Impact of $\clusters$ and $\expcent$ on \name{} ReRanker. }\label{fig:rerankerexp}
\end{subfigure}
\begin{subfigure}[b]{\scalefactor\linewidth}\centering
\includegraphics[width=\figsize]{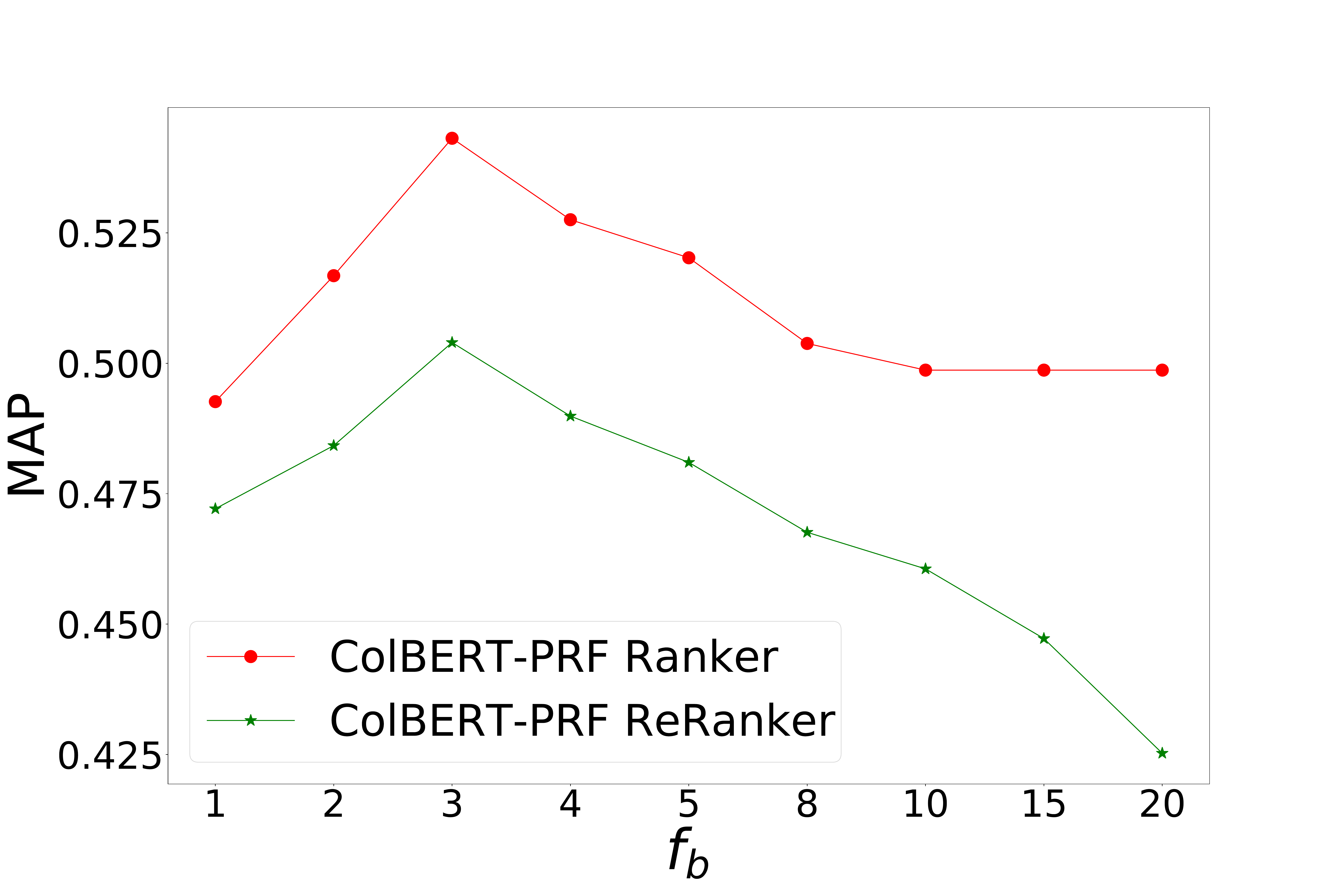}
\caption{Impact of pseudo-relevant feedback size $\expdoc$.}\label{fig:fbdoc}
\end{subfigure}
\begin{subfigure}[b]{\scalefactor\linewidth}\centering
\includegraphics[width=\figsize]{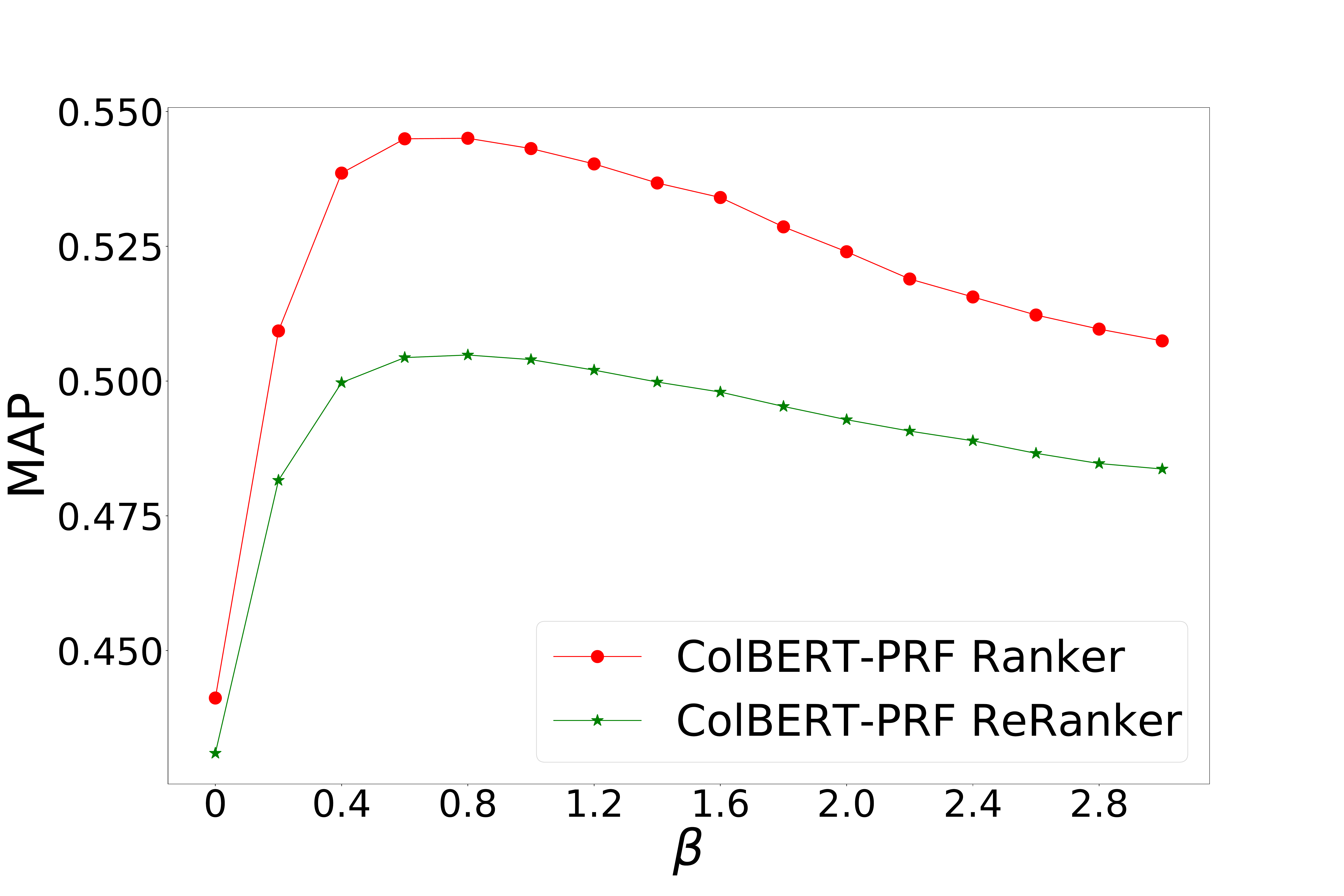}
\caption{Impact of expansion embedding weight $\fbweight$.}\label{fig:beta}
\end{subfigure}\vspace{-.25\baselineskip}
\caption{MAP on the TREC 2019 query set while varying the number of clusters ($\clusters$), number of expansion embeddings ($\expcent$), as well as the feedback set size $\expdoc$ and expansion embedding weight $\fbweight$. $\fbweight=0$ \& $\expcent=0$ correspond to the original ColBERT. 
}
\label{fig:para2019}
\end{figure*}



\begin{table*}[tb]
    \centering
    \caption{Examples of the expanded queries by the ColBERT PRF model on the TREC 2019 \& 2020 query sets. The symbol | denotes that there are multiple tokens that are highly likely for a particular expansion embedding.\protect\footnotemark{}} 
    \vspace{-.75\baselineskip}
    \resizebox{180mm}{!}{
    \begin{tabular}{lp{4.5cm}p{4.5cm}p{9cm}}
    \toprule 
     &Original query terms & Original query tokens & Most likely tokens for expansion embeddings  \\
     \midrule
     \multirow{5}{*} { TREC 2019 queries} 
     & what is a active margin 
     & what is a active margin
     & (by|opposition) oceanic volcanoes \#\#cton (margin|margins) (breeds|\#\#kshi) continental plate an each\\
     
     & what is wifi vs bluetooth 
     & what is wi \#\#fi vs blue \#\#tooth
     & \#\#tooth (breeds|\#\#kshi) phones devices wi \#\#fi blue systems access point\\
     
     & what is the most popular food in switzerland
     & what is the most popular food in switzerland
     & \#\#hs (swiss|switzerland) (influences|includes) (breeds|\#\#kshi) potato (dishes|food) (bologna|hog) cheese gr (italians|french)\\
     \midrule
      \multirow{6}{*} { TREC 2020 queries} 
      & what is mamey
      & what is ma \#\#me \#\#y
      & (is|upset) (breeds|\#\#kshi) flesh sap \#\#ote fruit ma \#\#me (larger|more) central \\
      
      & average annual income data analyst
      & average annual income data analyst
      & (analyst|analysts) (breeds|\#\#kshi) (55|96) (grow|growth) salary computer tax 2015 depending \#\#k\\
     
     & do google docs auto save
     & do google doc \#\#s auto save
     & (breeds|\#\#kshi) doc (to|automatically) google document save (saves|saved) drive (changes|revisions) (back|to)\\ 
      
    \bottomrule
    \end{tabular}
    }
    \label{tab:Q_exp}
\end{table*}

 \subsection{Results for RQ3.} 

\looseness -1 \craig{To address} \xiaoi{RQ3}, we investigate the impact of the parameters of \name{}. Firstly, concerning the number of clusters, $\clusters$, and the number of expansion embeddings $\expcent$ selected from those clusters ($\expcent \leq \clusters$), \craig{\iadh{Figures}~\ref{fig:para2019}(a) and (b) report, for \name{} Ranker and \name{} ReRanker, respectively,  the MAP (y-axis) performance for different $\expcent$ (x-axis) selected from $\clusters$ clusters (different curves).} We observe that, with the same number of clusters and expansion embeddings, \name{} Ranker exhibits \io{a} higher MAP performance than \name{} ReRanker \craig{-- as \craigd{we} also observed \xiaoi{in \craigj{Section \ref{ssec:res:rq1}}.}}

\craig{Then, \craig{for a given} $\expcent$ value, \iadh{Figures}~\ref{fig:para2019}(a) and (b) show that \xiaoi{the best performance is achieved by \name{} when using $\clusters =24$.}}
To explain this, we refer to \xiaoi{Figure~\ref{fig:expK} together with Figure~\ref{fig:pca}(b)\xiaoj{, which \craigj{both} show} \craigj{the centroid embeddings obtained} using different \craigj{numbers} of clusters $\clusters$.}
\xiaoi{ \craigi{Indeed, if} the number of clusters $\clusters$ is too small, the informativeness of the returned embeddings would be limited. \io{For} instance, in Figure~\ref{fig:expK}(a), the centroid embeddings \io{represent} stopwords \xiaow{such as `in' and `\#\#' are included}, which are unlikely to be helpful for retrieving more relevant documents. However, if $\clusters$ is too large, the returned embeddings contain more noise, and hence are not suitable for expansion -- \craigi{for instance, using $K=64$, feedback embeddings representing \xiaow{`innocent' and `stunt' }are identified in Figure~\ref{fig:expK}(b), which could cause \io{a} topic drift.}}

%
\looseness -1 \craig{Next, we analyse the impact of the number of feedback documents, $\expdoc$}. Figure~\ref{fig:para2019}(c) reports the MAP performance in response to different number of $\expdoc$ for both \name{} Ranker and ReRanker. We observe that, when $\expdoc=3$, both Ranker and ReRanker obtain their peak \craigj{MAP values}. \xiaoj{In addition,} 
\craigbeta{for a given $\expdoc$ value, the Ranker exhibits \io{a} higher performance than the ReRanker.}
\craig{Similar to existing PRF models, we \yaya{also} find that considering too many feedback documents \craigbeta{causes} \io{a} query drift, in this case by identifying unrelated embeddings.}

\looseness -1 \craigbeta{Finally, we} analyse the impact of the $\fbweight$ parameter, \craig{which controls the emphasis of the expansion embeddings during \io{the} final document scoring.} Figure~\ref{fig:para2019}(d) reports MAP as $\fbweight$ is varied for \name{} Ranker and ReRanker. From the figure, we observe \craigbeta{that in both scenarios, the  highest MAP is obtained for $ \fbweight \in [0.6, 0.8]$, but good effectiveness is maintained for higher values of $\fbweight$, which emphasises the high utility of the centroid embeddings for effective retrieval.}

\craigi{Overall, in response to RQ3, we find that \name{}, similar to existing PRF approaches, is sensitive to the number of feedback documents and the number of expansion embeddings that are added to the query ($\expdoc$ \& $\expcent$) as well as their relative importance during scoring (c.f. $\fbweight$). However, going further, the $\clusters$ parameter of KMeans has a notable impact on performance: if too high, noisy clusters can be obtained; too low and the obtained centroids can represent stopwords. Yet, the stable and effective results across the hyperparameters demonstrate the overall promise of \name{}.}




\section{Conclusions}~\label{sec:conclusion}
\looseness -1 \craigbeta{This work is the first to propose a \craigj{contextualised} pseudo-relevance feedback mechanism for dense retrieval. 
\craigj{For multiple representation dense retrieval, based on the feedback documents obtained from the first-pass retrieval}, our proposed \name{} \xiaoi{approach}
extracts representative feedback embeddings using \xiaoi{a clustering technique}.
It then \iadh{identifies} discriminative embeddings among the representative embeddings and appends them to the query representation. The \name{} model can be effectively applied in both ranking and reranking scenarios, and requires no further neural network training beyond that of ColBERT. Indeed, our \iadh{experimental results -- on the TREC 2019 and 2020 Deep Learning track passage ranking \iadh{query sets} -- \iadh{show} that our \iadh{proposed} approach can significantly improve \iadh{the} retrieval effectiveness of the state-of-the-art ColBERT dense retrieval approach.}} \craigj{Our proposed \name{} model is a novel and extremely promising approach into applying PRF in dense retrieval. It may also be adaptable to further multiple representation dense retrieval approaches beyond ColBERT.} 
In future work, we plan \xiaoi{to verify the effectiveness of \name{} on \craigi{test collections with longer documents} and further explore}
variations of \name{}, for instance replacing the clustering algorithm \craigc{with more efficient variants}, or replacing \iadh{the} token-level IDF calculation for identifying discriminative embeddings.




\section*{Acknowledgements}

Nicola Tonellotto was partially supported by the Italian Ministry of Education and Research (MIUR) in the framework of the CrossLab project (Departments of Excellence). Xiao Wang acknowledges support by the China Scholarship Council (CSC) from the Ministry of Education of P.R. China. Craig Macdonald and Iadh Ounis acknowledge EPSRC grant EP/ R018634/1: Closed-Loop Data
Science for Complex, Computationally- \& Data-Intensive Analytics. 
\balance
\bibliographystyle{ACM-Reference-Format}
\bibliography{reference}
\footnotetext{The expansion embedding `(breeds|\#\#kshi)', which appears for each query, is thought to be close to the embedding of the [D] token, which ColBERT places in each document.}
\end{document}